\documentclass[prd,reprint,aps,amsmath,amssymb,tightenlines,showpacs,nofootinbib]{revtex4-1}

\usepackage{graphicx}
\usepackage{dcolumn}
\usepackage{bm}
\usepackage[dvipsnames]{xcolor}
\usepackage[colorlinks=true,citecolor=blue,linkcolor=blue]{hyperref}

\newcommand\myshade{85}
\colorlet{mylinkcolor}{PineGreen}
\colorlet{mycitecolor}{BrickRed}
\colorlet{myurlcolor}{violet}
\hypersetup{
  linkcolor  = mylinkcolor!\myshade!black,
  citecolor  = mycitecolor!\myshade!black,
  urlcolor   = myurlcolor!\myshade!black,
  colorlinks = true
}

\def\beq{\begin{equation}\begin{aligned}}
\def\eeq{\end{aligned}\end{equation}}

\newcommand{\be}{\begin{equation}} 
\newcommand{\ee}{\end{equation}}

\newcommand{\dd}{\mathrm{d}}

\newcommand{\Msun}{\ensuremath{\mathrm{M}_\odot}}

\def\gev{\, {\rm GeV}}

\def\ev{\, {\rm eV}}

\begin{document}

\title{Superradiance and the Spins of Black Holes from LIGO and X-ray binaries}

\author{Nicolas~Fernandez}
\email{nicofer@illinois.edu}
\affiliation{Department of Physics, University of Illinois at Urbana-Champaign, Urbana, IL 61801, USA}
\affiliation{Department of Physics, 1156 High St., University of California Santa Cruz, Santa Cruz, CA 95064, USA}
\affiliation{Santa Cruz Institute for Particle Physics, 1156 High St., Santa Cruz, CA 95064, USA}

\author{Akshay~Ghalsasi}
\email{aghalsas@ucsc.edu}
\affiliation{Department of Physics, 1156 High St., University of California Santa Cruz, Santa Cruz, CA 95064, USA}
\affiliation{Santa Cruz Institute for Particle Physics, 1156 High St., Santa Cruz, CA 95064, USA}

\author{Stefano~Profumo}
\email{profumo@ucsc.edu}
\affiliation{Department of Physics, 1156 High St., University of California Santa Cruz, Santa Cruz, CA 95064, USA}
\affiliation{Santa Cruz Institute for Particle Physics, 1156 High St., Santa Cruz, CA 95064, USA}


\begin{abstract}
Measurements of the spin of stellar mass black holes (BHs) are now possible both through LIGO observations of binary BH mergers and for BHs in X-ray binary systems. The spins of BHs as inferred from LIGO observations suggest that BH spins are on the lower end of what is expected for a ``flat'' distribution of spins, while those from BHs in X-ray binaries tend to be large. Superradiance, a process that can effectively reduce the spin of BHs before they merge, could explain the lower observed spins in binary BH mergers for a non self-interacting light boson. In this paper, we use Bayesian analysis to infer the posterior probability distribution for the mass of a light boson that could fit LIGO data. We also analyze spins of BHs from X-ray binaries, and find that the X-ray binary data can be explained by superradiance due to a light boson with large self-interactions. We infer the mass range for such a boson that is consistent with the X-ray binary data.
\end{abstract}

\maketitle

\section{Introduction}
\label{sec:intro}
Spinning black holes (``Kerr'' black holes, BHs) possess a region outside the event horizon  where matter cannot remain in rest relative to an external observer due to the effect of inertial frame dragging. This region is known as the ``{\em ergoregion}''. Any  plane wave (matter wave or light wave) infalling onto the BH and crossing the ergoregion co-rotates with the BH as it gets scattered, effectively extracting angular momentum from the BH. 
This is known as the {\it Penrose superradiance process} \cite{Penrose:1969pc,Teukolsky:1974yv}. Massive bosons can form hydrogen-like bound states around the BH and the bound state wavefunction has an overlap with the ergoregion. This allows for continually extracting angular momentum and energy from the BH by exponentially populating the bound states \cite{Arvanitaki_2010,Arvanitaki_2011,baumann2019spectra,PhysRevD.22.2323,Brito_2015,Brito_2015_1,Witek_2013,Dolan_2007}. Ultralight bosons in the mass  range of $10^{-11}{\rm eV} - 10^{-13}{\rm eV} $  impact BH spins with masses in the range $\mathcal{O}(1 \Msun) - \mathcal{O}(100 \Msun)$ where $\Msun$ is the mass of the sun. Measurements of BH spins in this mass range can thus be used to constrain or argue for the presence of ultralight bosons.

The advent of the Laser Interferometer Gravitational-Wave Observatory (LIGO)\cite{Abbott:2016blz} has ushered a new era of observational astronomy and cosmology. LIGO data allow to infer the mass and spin of  merging BHs; LIGO is expected to collect $\mathcal{O}(1000)$ events through its lifetime in the mass range of $\mathcal{O}(1 \Msun) - \mathcal{O}(100 \Msun)$. Using LIGO data to constrain ultralight bosons has been considered before. In \cite{Stott_2018} they consider the six BH spin measurements from the LIGO O1 run as part of their analysis. However their analysis is not Bayesian in nature. In \cite{Arvanitaki_2017} the authors project the number of LIGO events needed to disentangle the effects of superradiance at $2\sigma$ confidence. Recently, \cite{Ng:2019jsx} performed a Bayesian analysis on the 20 O2 LIGO events.  Our analysis is consistent with their findings. However, we go beyond: for instance, we also consider the effects on superradiance by the gravitational perturbation caused by the companion BH in the binary system. 

Another source of measurement of BH mass and spins are X-ray binary systems in which a star and a BH are in a binary system with the star providing the source material for the accretion disk around the BH which is visible in X-ray emissions. X-ray binaries can have very large spins, as evident from Table \ref{table:parameters_xrays}, and can be powerful in constraining ultralight boson masses. In fact, several papers have already used X-ray binaries to constrain ultralight boson masses \cite{Cardoso_2018,Stott_2018,Arvanitaki_2015,sun2019search}. In \cite{Stott_2018}, they use X-ray binary measurements to constrain non-interacting boson masses between $7\times 10^{-14}{\rm eV} - 2\times 10^{-11}{\rm eV}$. However, they do not consider the effects on superradiance by the gravitational perturbation caused by the companion star. In \cite{Arvanitaki_2015}, they consider the effects from gravitational perturbation caused by the companion star as well as the effect of self-interactions to constrain boson masses and self-interactions. In their exclusions, they point out the possibility of the spins of the BHs being stuck on Regge trajectories. Our analysis uses BH spins and masses, somewhat different than used by \cite{Arvanitaki_2015} and we find the allowed mass range for ultralight bosons where this possibility is realized.

Our paper is organized as follows. In Section \ref{sec:superrad},  we briefly describe the process of superradiance for an isolated BH as well as for a BH in the presence of gravitational perturbation caused by a companion. In Section \ref{sec:LIGO}, we present our Bayesian analysis and derive posteriors for the ultralight boson masses using LIGO data.  In Section \ref{sec:xray} we analyze the X-ray binary data and derive the allowed mass range and self-interaction of the ultralight bosons that explain the data. We conclude in Section \ref{sec:conclusion}.

\section{Superradiance and Regge Plots}
\label{sec:superrad}
 Superradiance can be very efficient at extracting angular momentum from BHs when the bosons bound by the gravitational potential have a reduced Compton wavelength $\lambda_{c}$ comparable to the BH's gravitational radius $R_{g}$. Thus, the gravitational fine{-}structure constant can be defined as 
\beq
\alpha \equiv \dfrac{R_{g}}{\lambda_{c}} = \dfrac{G_{N} M \mu}{\hbar c} \,
\eeq
where $M$ and $\mu$ are the mass of the BH and the boson, respectively. Notice that here we use natural units where $\hbar=c=1$.

However, superradiance cannot spin down a BH to arbitrarily low spins.  The extraction of spin from the BH continues until the so-called superradiance condition is satisfied. The superradiance condition is a consequence of the  second law of thermodynamics and can be derived by using the fact that the entropy (area of the event horizon) has to increase \cite{Arvanitaki_2010}. For a boson with wave form parametrized by $\Psi \propto e^{-i\omega t + i m \phi}\psi(r,\theta)$, the superradiance condition is given by
\begin{align}
\label{eq:supercondition}
0 < \omega \approx \mu < m \omega_{+} \, ,
\end{align}
where $\omega_{+} = \frac{\chi}{2R_{g}(1+\sqrt{1-\chi^{2}})}$ can be thought of as the angular velocity of the BH and $\chi = \frac{J}{M R_{g}}$ is the spin of the BH with angular momentum $J$. We have ignored $\mathcal{O}(\alpha^{2})$ corrections to $\omega$.

For $\alpha = \mu R_{g} \ll l$, the rate of superradiance for each bound state can be approximated analytically\footnote{A better approximation of the rates can be done semi-analytically as has been done in \cite{Arvanitaki_2011,Stott_2018}. However, we will use the analytic approximation here.}. The rate of growth of a superradiant bound state with radial, angular and magnetic quantum numbers ($n,l,m$) has been calculated \cite{Arvanitaki_2011}  and are given by
\begin{align}
\label{eq:rate}
\Gamma_{nlm} &= 2\mu \alpha^{4l+4} r_{+}(m\omega_{+}-\mu)C_{lmn}\\
C_{nlm} &= \frac{2^{4l+2}(2l+n+1)!}{(l+n+1)^{2l+4}n!}\left(\frac{l!}{(2l)!(2l+1)!}\right)^{2} \times\\
&\prod_{j=1}^{l}\left(j^{2}\left(1-\chi^{2}\right)+4r^{2}_{+}(m\omega_{+}-\mu)^{2}\right)
\end{align}
where $r_{+} = R_{g}(1+\sqrt{1-\chi^{2}})$.

\begin{figure}[t]
\centering
\includegraphics[width=0.48\textwidth]{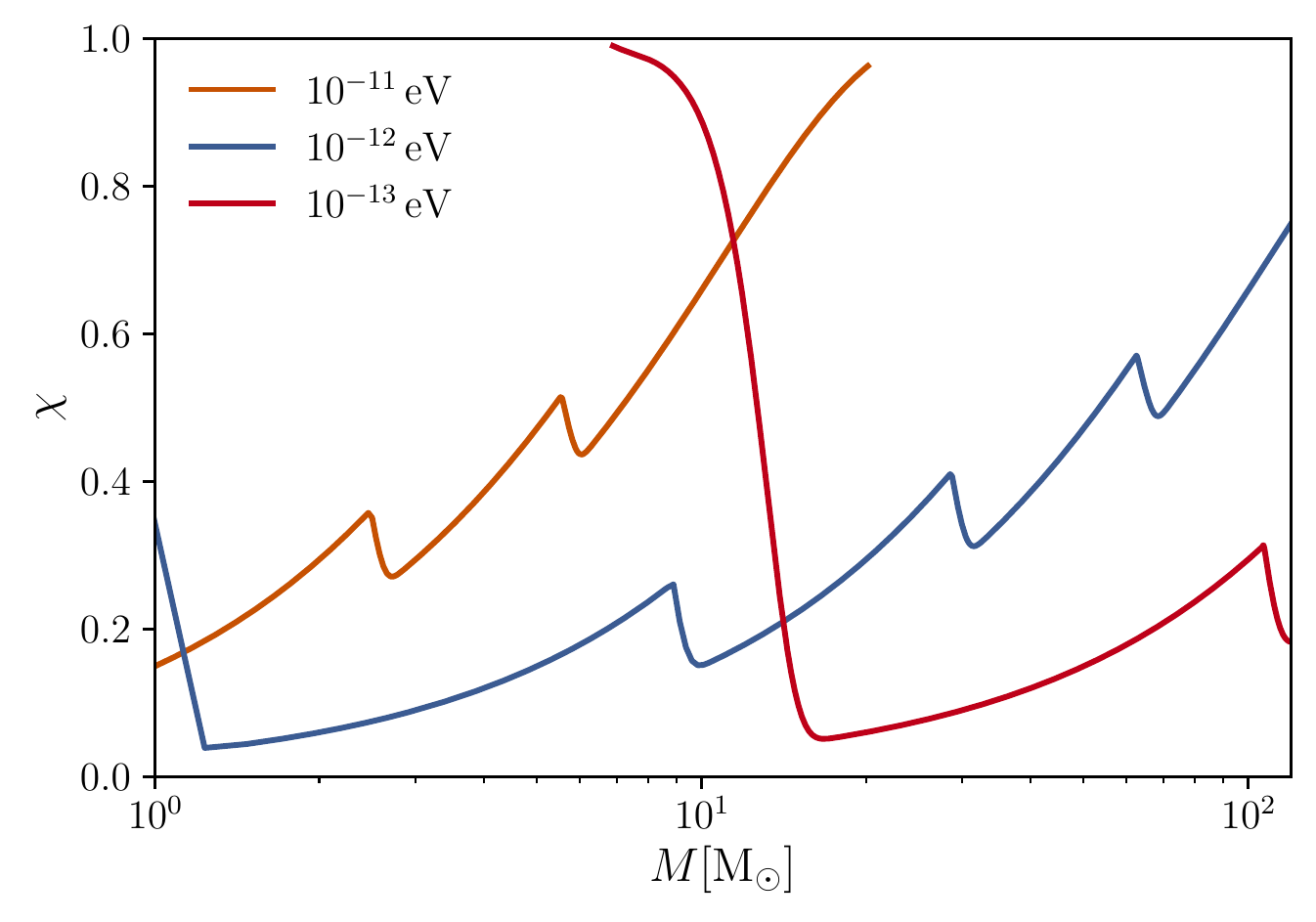}
\caption{Regge Plots assuming superradiance is effective for $10\, \rm Gyr$. We have only shown up to the first 4 peaks since we consider those for our analysis. }
\label{fig:ReggePlots}
\end{figure}

In the limit where the self-interactions of the light bosons, or the perturbation due to the gravitational field of a companion object, are not strong to significantly perturb the superradiant wavefunctions, the growth of the number of bosons in any superradiant state can be written as $$\frac{dN}{dt} = \Gamma_{nlm}N \, ,$$ where $N$ is the number of bosons in the superradiant state. Note that in order to extract sufficient spin from the system, $\Delta \chi \approx \mathcal{O}(1)$, one needs to extract angular momentum (and hence have occupation levels) of order $$\frac{M^{2}}{M^{2}_{\rm pl}}\simeq 10^{76}\left(\frac{M}{M{\odot}}\right)^{2}\, ,$$ where $M_{\rm pl}$ is the Planck mass. The number of e-foldings required to reach such a high occupation number is $\approx 180$. Thus, as long as the timescale of the system $$\tau_{\rm sys} \geq \frac{180}{\Gamma_{nlm}}\, ,$$ superradiance will reduce the spin of the BH until the superradiance condition in Eq. \ref{eq:supercondition} is no longer satisfied, and superradiance shuts off. We can use this to construct exclusion plots as a function of BH mass and spin known as Regge plots for a given boson mass and the time associated with the system. The Regge exclusion plots for LIGO BHs for certain boson masses and timescale corresponding to $10\, \rm Gyr$ are shown in Fig. \ref{fig:ReggePlots}.

Each Regge trajectory has multiple peaks. This happens because after the $l=m$ level saturates the superradiant condition (Eq. (\ref{eq:supercondition})), the $l+1=m+1$ level takes over. Here, we have only shown the first four peaks of the superradiance process and we will consider those in our analysis\footnote{Including higher peaks does not affect our results significantly and for $l > 5$ the superradiance timescales are larger than the age of the universe.}.

In the presence of a companion object (a BH or a star), the gravitational perturbation from the companion object will cause the superradiance levels to mix, causing superradiance to shut down \cite{Arvanitaki_2011,Arvanitaki_2015}. BHs in X-ray binaries have a companion star whose gravitational perturbation can cause level mixing in the superradiance levels and, as a result, cause the superradiance to shut down\footnote{The gravitational perturbation due to the accretion disk is subdominant \cite{Arvanitaki_2015}.}. For BH binary mergers, if the BHs are formed in a binary, their superradiance evolution is affected by the companion. The gravitational perturbation due to a companion of mass $M$ is given by
\beq
\delta V = \mu \left(\frac{M}{M_{BH}}\right) \frac{R_{g}}{R}\left(1+Y_{1,m}\frac{r_{c}}{R}+...\right) \, ,
\eeq
where $r_{c} = \frac{(n+l+1)^{2}}{\alpha^{2}}R_{g}$ is the boson cloud radius and $R$ is the distance from the companion to the BH. In presence of such a gravitational perturbation $\delta V$, the perturbed state is given by
\beq
|\psi\rangle = N\left(|\psi^{nlm}\rangle + \frac{\langle \psi^{nlm}|\delta V| \psi^{n'l'm'}\rangle}{\Delta E} |\psi^{n'l'm'}\rangle \right), 
\eeq
where $|\psi^{n,l,m}\rangle$ is the superradiant state, $|\psi^{n',l',m'}\rangle$ is the dumped state, $\Delta E = \left(\frac{\alpha}{l}\right)^{4}\mu$ is the energy difference for $n' = n, l' \neq l$ and $N$ is the normalization of the state. The superradiant state extracts spin and energy from the BH while the dumped state dumps them back in. Thus, for superradiance to be active, the net flux crossing the BH horizon needs to be positive, which translates into the following condition:
\beq
\label{eq:srlevelmix}
\left|\frac{\Gamma^{n'l'm'}}{\Gamma^{nlm}}\right| \left| \frac{\langle \psi^{nlm}|\delta V| \psi^{n'l'm'}\rangle}{\Delta E}\right|^{2}  < 1 \, .
\eeq

The superradiance levels that are the most active for us have $n=0,l=m$. The level mixing between superradiant and dumping levels is the strongest for $l'=m'=l-1$. Then using Eq. (\ref{eq:srlevelmix}), the condition we get for superradiance to be active is
\beq
\label{eq:suppercondgrav}
\left|\frac{\Gamma^{0,l-1,m-1}}{\Gamma^{0,l,m}}\right|\left(\frac{l}{\alpha}\right)^{8}\left(\frac{n+l+1}{\alpha}\right)^{4}&\\
\times \left(\frac{R_{g}}{R}\right)^{4}\langle \psi^{0,l,m}|Y_{1,1}| \psi^{0,l-1,m-1}\rangle^{2} &< 1 \, .
\eeq
The only unknown quantity in the above condition is $\frac{R_{g}}{R}$. If the LIGO BHs are formed in a binary, then the lifetime of the binary is controlled by the inspiral lifetime which can be related to the initial binary separation $R$. Assuming equal mass binaries\footnote{Each LIGO binary has mass ratio $q\neq 1$ and hence the superradiance curves should be calculated separately for each LIGO event. However, in practice the superradiance curves don't change much if we assume equal mass binaries because of the strong dependence of $\alpha$ in the superradiance condition in Eq. (\ref{eq:suppercondgrav}). } and the rate of luminosity of GW wave emission given in \cite{2016}, we can write
\beq
\label{eq:rratio}
\frac{R_{g}}{R} = 3.5\times10^{-6}\left(\frac{10^{10} \, \rm yr}{\tau_{binary}}\right)^{1/4}\left(\frac{M_{BH}}{30 \Msun}\right)^{1/4}.
\eeq

The spin down of a BH in presence of self-interactions will be discussed briefly in Sec. \ref{sec:xray}. See \cite{Arvanitaki_2011,Arvanitaki_2015} for a more thorough discussion. For other interesting effects and observables that have been considered in the literature for superradiance for BHs that are part of binaries see, \cite{Baumann_2019,Berti_2019,zhang2019dynamic}. 
\section{LIGO Events}
\label{sec:LIGO}

\begin{table}[t]
  \centering
\scalebox{0.9}{
\begin{tabular}{ l r r r r r r r r}
\hline 
Event & & \multicolumn{1}{c}{$M_1 [\Msun]$}    & &   \multicolumn{1}{c}{$M_2 \left[ \Msun \right]$}  & & \multicolumn{1}{c}{$\chi_{1}$} & & \multicolumn{1}{c}{$\chi_{2}$}  \\  [0.2ex]  \hline   \hline  \\ [-1.4ex]
GW150914  & & $35.6^{+ 4.7}_{-3.1}$   & & $30.6^{+3.0}_{-4.4}$  & & $0.28^{+0.57}_{-0.25}$ & & $0.34^{+0.53}_{-0.30}$\\ [0.14cm]
GW151012  & & $23.2^{+ 14.9}_{-5.5}$  & & $13.6^{+4.1}_{-4.8}$  & &  $0.33^{+0.54}_{-0.29}$ & & $0.45^{+0.48}_{-0.40}$\\ [0.14cm]
GW151226  & & $13.7^{+ 8.8}_{-3.2}$   & & $7.7^{+2.2}_{-2.5}$   & &  $0.57^{+0.36}_{-0.43}$ & & $0.51^{+0.44}_{-0.45}$\\ [0.14cm]
GW170104  & & $30.8^{+ 7.3}_{-5.6}$   & & $20.0^{+4.9}_{-4.6}$  & &  $0.34^{+0.52}_{-0.30}$ & & $0.43^{+0.48}_{-0.38}$\\ [0.14cm]
GW170608  & & $11.0^{+5.5}_{-1.7}$    & & $7.6^{+1.4}_{-2.2}$   & &  $0.32^{+0.50}_{-0.28}$ & & $0.40^{+0.52}_{-0.36}$\\ [0.14cm]
GW170729  & & $50.2^{+16.2}_{-10.2}$  & & $34.0^{+9.1}_{-10.1}$ & &  $0.69^{+0.28}_{-0.55}$ & & $0.55^{+0.40}_{-0.49}$\\ [0.14cm]
GW170809  & & $35.0^{+8.3}_{-5.9}$    & & $23.8^{+5.1}_{-5.2}$  & &  $0.32^{+0.53}_{-0.29}$ & & $0.42^{+0.50}_{-0.37}$\\ [0.14cm]
GW170814  & & $30.6^{+ 5.6}_{-3.0}$   & & $25.2^{+2.8}_{-4.0}$  & &  $0.40^{+0.52}_{-0.36}$ & & $0.42^{+0.51}_{-0.37}$\\ [0.14cm]
GW170818  & & $35.4^{+7.5}_{-4.7}$    & & $26.7^{+4.3}_{-5.2}$  & &  $0.46^{+0.48}_{-0.41}$ & & $0.46^{+0.47}_{-0.42}$\\ [0.14cm]
GW170823  & & $39.5^{+ 11.2}_{-6.7}$  & & $29.0^{+6.7}_{-7.8}$  & &  $0.42^{+0.49}_{-0.37}$ & & $0.45^{+0.48}_{-0.40}$\\ [1ex] \hline
\end{tabular}
}
\caption {Mass and dimensionless spin magnitude for the ten BBH mergers events detected during LIGO's O1 and O2 runs. The parameters are median values, with 90\% confidence intervals \cite{LIGOScientific:2018mvr}.}
\label{table: parameters}
\end{table}

\begin{figure}[htp]
\centering
\includegraphics[width=0.48\textwidth]{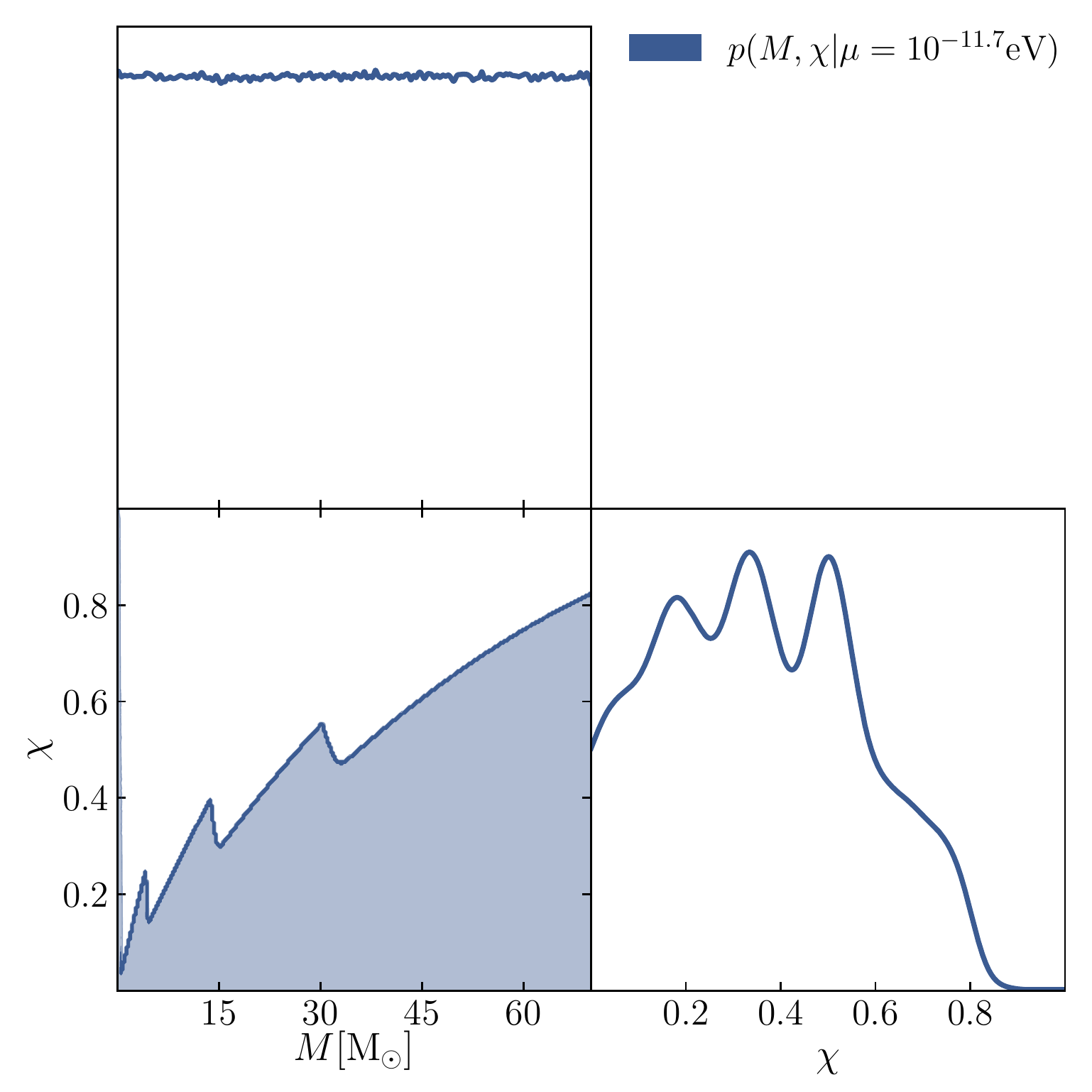}
\caption{Probability density distribution $p \left(  m,\chi \mid  \mu \right)$ for  $\mu=10^{-11.7}\ev$ assuming the initial spin distribution to be flat.}
\label{fig:Distribution_m_chi}
\end{figure}

The existence of ultralight bosons would lead to a clear signature in the BH spin{-}mass plane in LIGO analysis. Unfortunately, the current measurements on the BHs spin magnitudes are poorly constrained since LIGO is only sensitive to the last few cycles before merger \cite{Vitale:2016avz}. The objective of this analysis is to calculate $p(\mu|\{d^{i}\})$, the posterior density function for the mass of the ultralight bosons $\mu$, given the data from the $10$ BH binary mergers  observed by LIGO shown in Table \ref{table: parameters} \footnote{We don't consider the recent claimed detections in \cite{Zackay_2019,venumadhav2019new,zackay2019detecting} for our analysis.}. 
\subsection{Analysis}
Before we explain our statistical analysis, we clarify the underlying assumptions we make for the calculation of the superradiance from the light boson. In presence of large self-interactions, the spin down of a BH can be significantly modified due to bosenovas and shut down of superradiance due to level-mixing \cite{Arvanitaki_2011,Yoshino_2012,Yoshino_2015}. The larger the self-interactions, the more ineffective superradiance is in spinning down the BH. However, with only 20 measured BH spins by LIGO, it will not be illuminating to add another dimension of self-interactions to our analysis. Hence, for our LIGO analysis, we assume that our light boson {\em does not have any self-interactions}. While it is true that non-interacting light bosons in the mass range we consider for our LIGO analysis have already been ruled out by X-ray binaries \cite{Stott_2018,Arvanitaki_2015}, one can formulate models to reconcile X-ray binary data with non-interacting bosons for LIGO analysis. One way to avoid X-ray binary constraints is to invoke a chameleon-like scalar which becomes massive in the presence of baryons in the accretion disk shutting down superradiance. Another possibility is a quintessence{-}like field whose mass is a function of time which is light in the early universe during LIGO mergers but becomes heavier later in the universe when X-ray binaries form.

For our analysis we assume that all the binaries have the same merger time $\tau_{\rm merger}$. Furthermore, we do a separate analysis for two different merger timescales i.e. the shortest timescale possible for stellar-mass binaries $\tau_{\rm merger} = 10^{7} \, \rm yr$ or the longest timescale $\tau_{\rm merger} = 10^{10} \, \rm yr$. \cite{Dominik:2013tma,Morscher_2015,TheLIGOScientific:2016htt}.  
We also assume that superradiance is effective throughout the merger process. We assume two scenarios for BH mergers. In the first scenario, our BHs were well separated before merger and hence spun down in isolation before merging. In the second scenario, we assume that the BHs start out in a binary hence the effects of superradiance on either BH is diminished due to the presence of gravitational perturbation from the companion \cite{Arvanitaki_2017}.  We name the first one isolated scenario and the second on the companion scenario. For calculating the effect of gravitational perturbation for the companion scenario, we assume both BHs are $30 \Msun$. This does not affect our results since the superradiance rates in presence of a companion are a mild function of BH masses (see Eqs. (\ref{eq:suppercondgrav}) and (\ref{eq:rratio})).

\begin{figure*}[htp]
\centering
\includegraphics[width=0.47\textwidth]{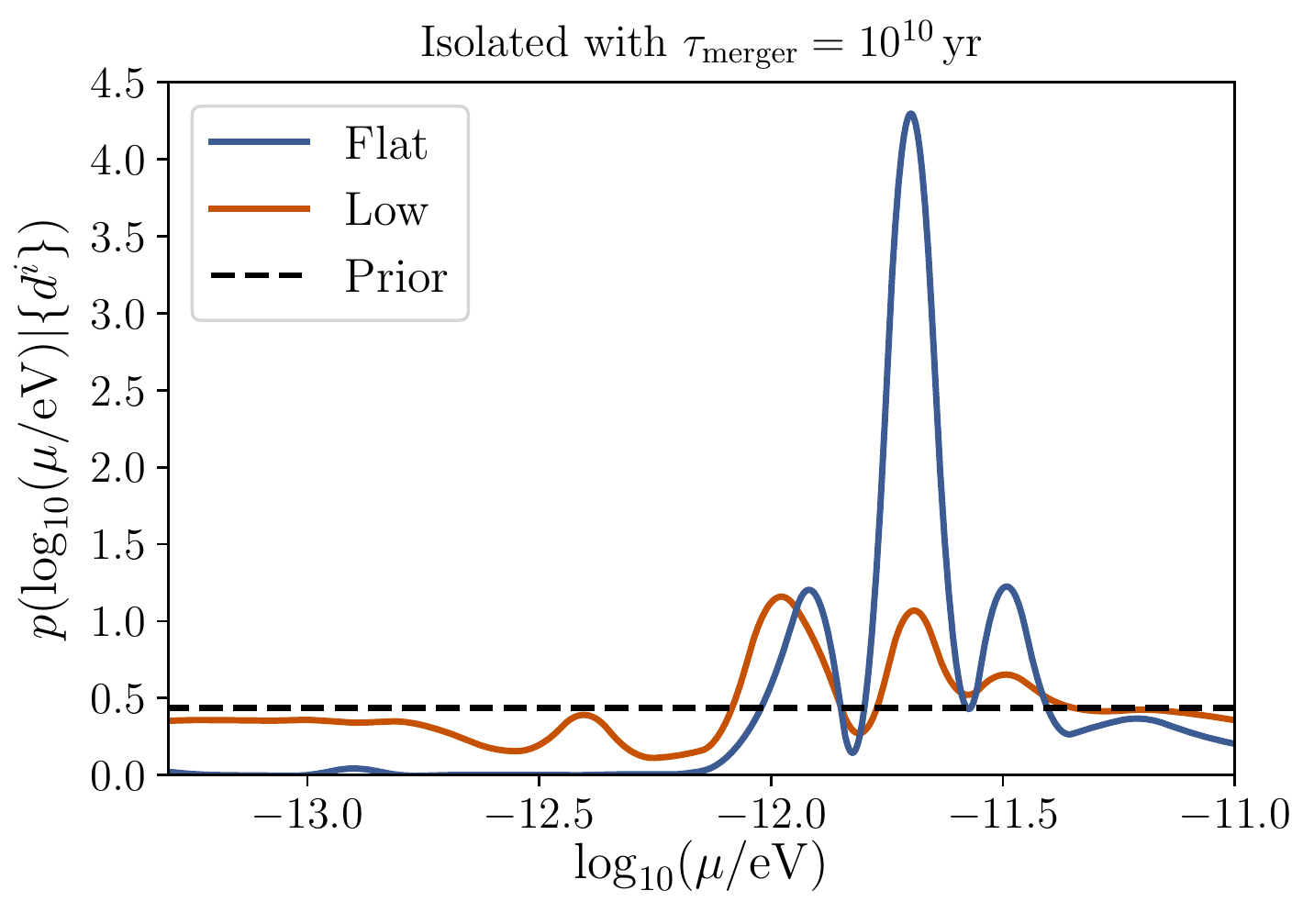}
\includegraphics[width=0.47\textwidth]{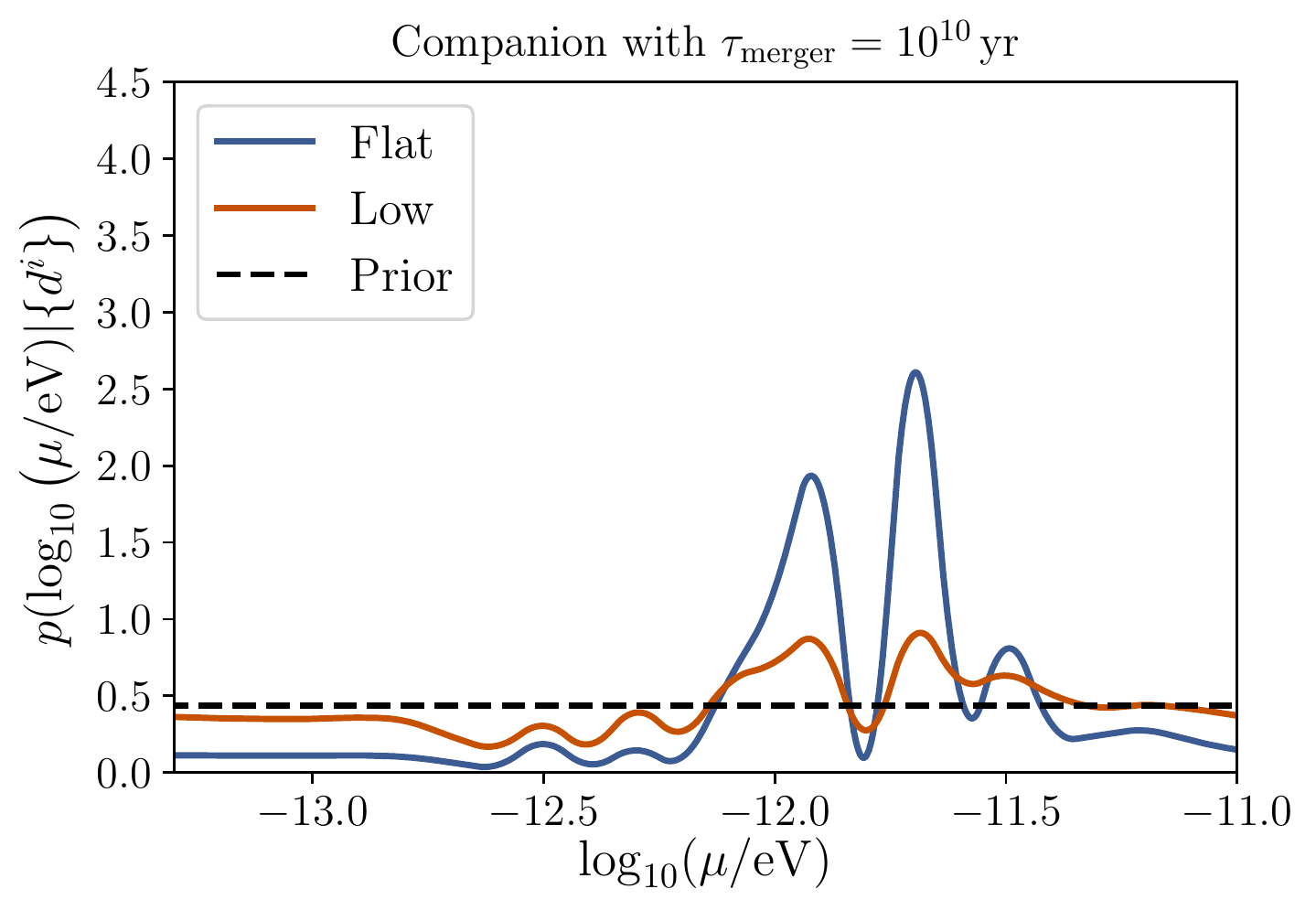}
\includegraphics[width=0.47\textwidth]{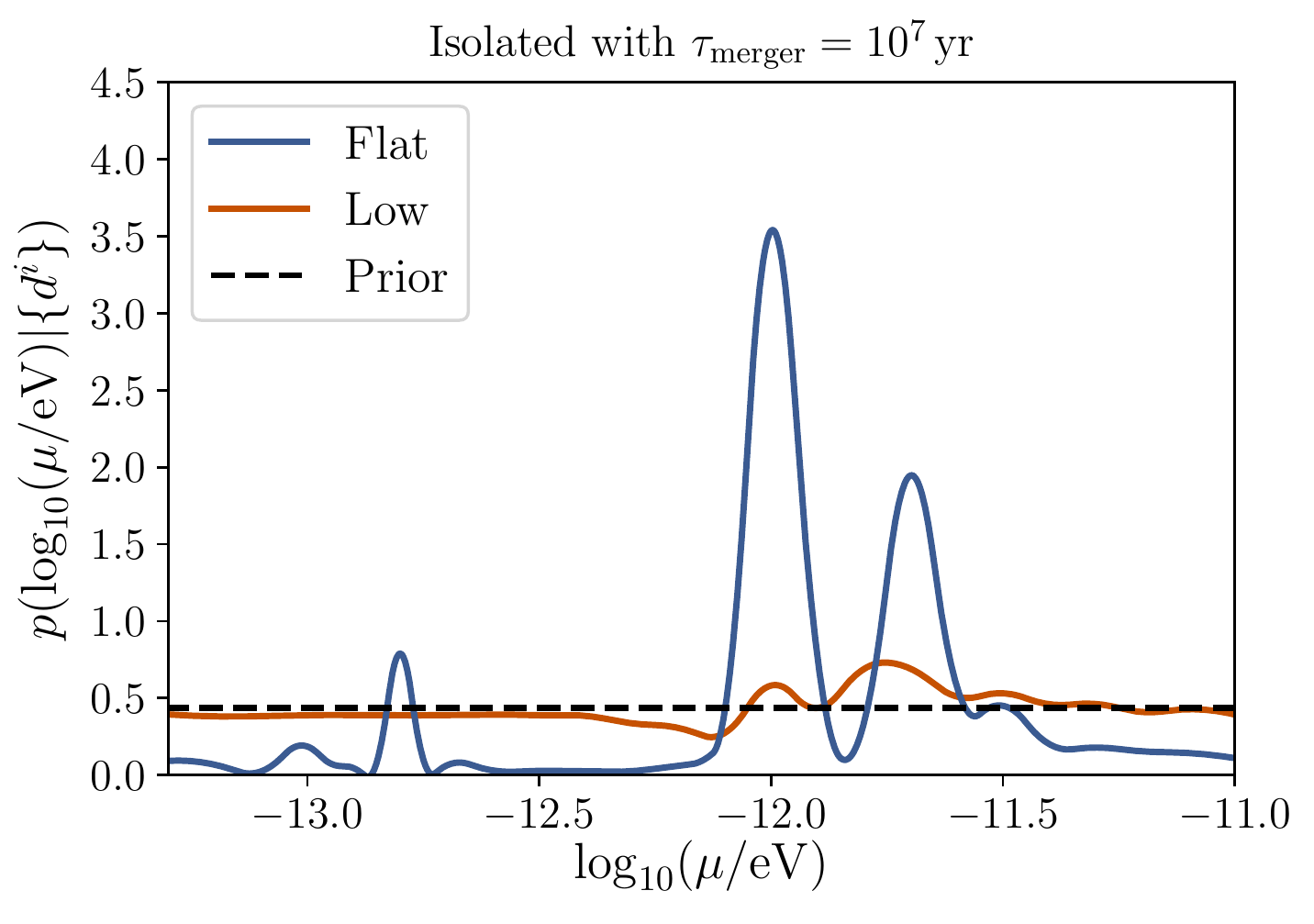}
\includegraphics[width=0.47\textwidth]{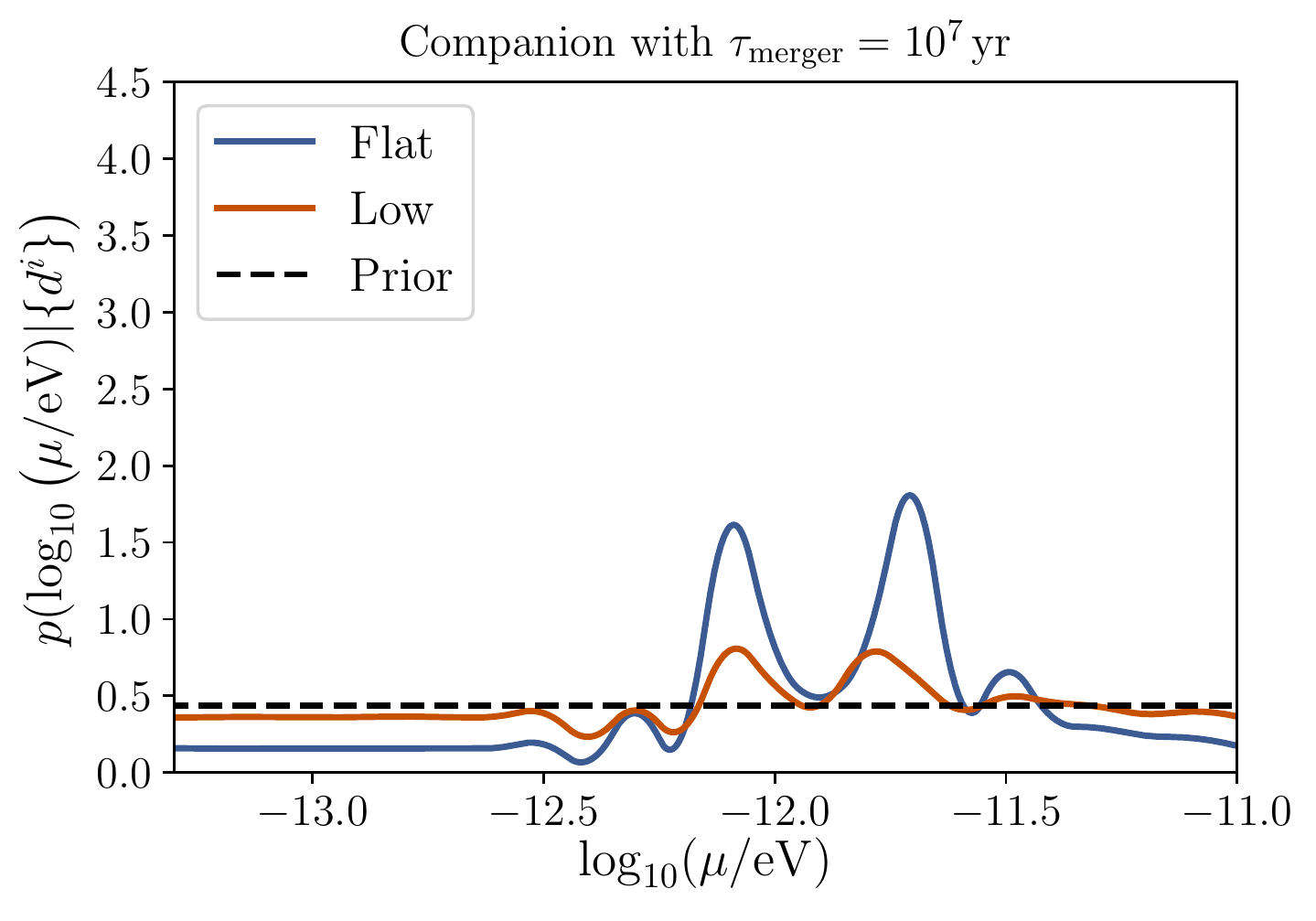}
\caption{Posterior distribution of the boson mass $\mu$ for the scenario where the BHs spin down in isolation before merging (\textit{left}) and for when 
the effects of superradiance is diminished due to the presence of the gravitational perturbation due to the companion during the merge (\textit{right}). For each scenario, we consider two cases for the initial spin magnitude distribution: low (blue) and flat (orange), while the black dashed line is the log-uniform distribution prior distribution assumed for $\mu$.}
\label{fig:Posterior_mass}
\end{figure*}

We utilize a {\em hierarchical Bayesian analysis} \cite{Mandel:2018mve} to derive the posterior for the light boson mass $\mu$. Assuming that the events are uncorrelated, the joint likelihood can be expressed as 
\beq
p\left( \{d^{i}\}  \mid  \Lambda \right) = \prod_{i=1}^{N_\mathrm{obs}}  p\left( d^{i}  \mid  \Lambda \right),
\eeq
where
\beq
p\left( d^{i}  \mid   \Lambda \right) = \int \dd \theta^i \, p \left(  d^{i}  \mid  \theta^i \right) p \left( \theta^i  \mid  \Lambda \right)
\eeq 
is the individual likelihood for the $i$th event from the measured data $d^{i}$ given a hyperparameter $\Lambda$. In our case, $\Lambda$ is just the mass of the ultralight boson $\mu$ and $\theta$ are the set of intrinsic parameters $(M_{1},M_{2},\chi_{1},\chi_{2})$. Here, $M_{1}$ and $M_{2}$ are the masses of the BHs such that $M_{1}>M_{2}$, and $\chi_{1}$ and $\chi_{2}$ are their respective spin magnitudes. 

Employing the joint likelihood for $N_{\mathrm{obs}}$ events, the posterior density function for the ultralight bosons mass $\mu$ can be expressed as follows:
\beq
p \left( \mu \mid \{ d^i \} \right) \propto & \left[  \prod_{i=1}^{N_\mathrm{obs}} \int p\left( d^i \mid M_{1}^{i},M_{2}^{i},\chi_{1}^{i},\chi_{2}^{i} \right) p\left(M_{1}^{i},M_{2}^{i} \right) \right. \\
 & \enspace \left. \prod_{j=1}^{2} p \left(  M_{j}^{i},\chi_{j}^{i} \mid  \mu \right) \dd M_{j}^i \, \dd \chi_{j}^i \right] p \left( \mu \right)\,,
 \label{eq:Posterior_mu}
\eeq
where $i$ labels the event and $j$ labels the individual BHs in the binary. Here, $p\left( d^i \mid M_{1}^{i},M_{2}^{i},\chi_{1}^{i},\chi_{2}^{i} \right)$ is the marginalized likelihood and $p(M_1,M_2)$ is  the mass prior for the BH masses which we take to be uniformly distributed. The term $p \left(  M_{j}^{i},\chi_{j}^{i} \mid  \mu \right)$ parametrizes the prior of $(M_{j}^{i},\chi_{j}^{i})$ in presence of superradiance for a given mass $\mu$ of the light boson. Our superradiance prior depends on the natal BHs spin distribution $p(\chi_{I})$\footnote{Note that superradiance extracts the mass of the BH as well. However we ignore this effect since  the change in mass is of $\mathcal{O}(10\%)$ for the BH masses we consider and is within the errors of the measurement.}. In this work, we consider two initial spin magnitude distributions:

\beq
\label{eq:spinprior}
p(\chi_{I})=1 &: \rm Flat\;spin\;distribution\\ 
p(\chi_{I})=2(1-\chi_{I}) &:\rm Low\;spin\;distribution
\eeq

In Fig.~\ref{fig:Distribution_m_chi}, we show an example of the mass-spin magnitude prior for a boson of mass $10^{-11.7}\ev$, assuming a flat spin magnitude distribution before superradiance takes place over $10^{10} \, \rm yr$. Therefore, if the initial spin $\chi_{I}$ lies within the superradiance exclusion region, the BH undergoes superradiance until the final BH spin $\chi_{s}$ settles on the lowest possible Regge trajectory available to it. On the other hand, if the initial spin lies outside of the superradiance region, the BH spin magnitude at merger is just the initial spin $\chi_{I}$. Thus, the spin magnitude distribution for a single BH can be written as 
\beq
p \left(  M,\chi \mid  \mu \right) = \left(1-N\right)p\left(\chi_{I}\right) + N \delta \left(\chi-\chi_{s}\left(M,\mu\right)\right) \,,
\nonumber 
\eeq
where $N$ is the fraction of BHs that have $\chi_{i}>\chi_{s}$ which depends on the initial spin magnitude distribution. For the low initial spin $N=(1-\chi_{s})^{2}$ and for a initial flat spin distribution, $N$ is just $(1-\chi_{s})$.
Finally, $p \left( \mu \right)$ is the prior distribution for the hyperparameter $\mu$. We employ a log-uniform distribution with range $10^{-13.3} \ev \leq \mu\leq 10^{-11} \ev $, which is the mass range relevant for LIGO.

We perform the integral in Eq. (\ref{eq:Posterior_mu}) by approximating it with a discrete sum over the LIGO posterior samples of the re-weighted priors. This allows us to evaluate the likelihood in terms of the LIGO posterior probability density without reanalyzing the original LIGO data\footnote{For e.g. see \cite{Farr:2017uvj,Talbot:2017yur,Farr:2017gtv,Tiwari:2018qch,Fernandez:2019kyb} where they follow a similar prescription.}. We would like to emphasize that the LIGO values from Table \ref{table: parameters} have been obtained by using uniform prior distributions in $(M_1,M_2)$, $\chi_1$, $\chi_2$ (with the constrain $M_{1} > M_{2}$). In the case of superradiance, these values will not longer hold true, because now the spin magnitude at merger will depend on the mass of the BH and the ultralight boson as seen in Fig.~\ref{fig:Distribution_m_chi}. This observation has been overlook in previous studies \cite{Stott_2018}. Thus, the values and the $90\%$ credible intervals of Table \ref{table: parameters} are subject to variations when different prior distributions are used \cite{Vitale:2017cfs}.

\begin{figure*}[htp]
\centering
\includegraphics{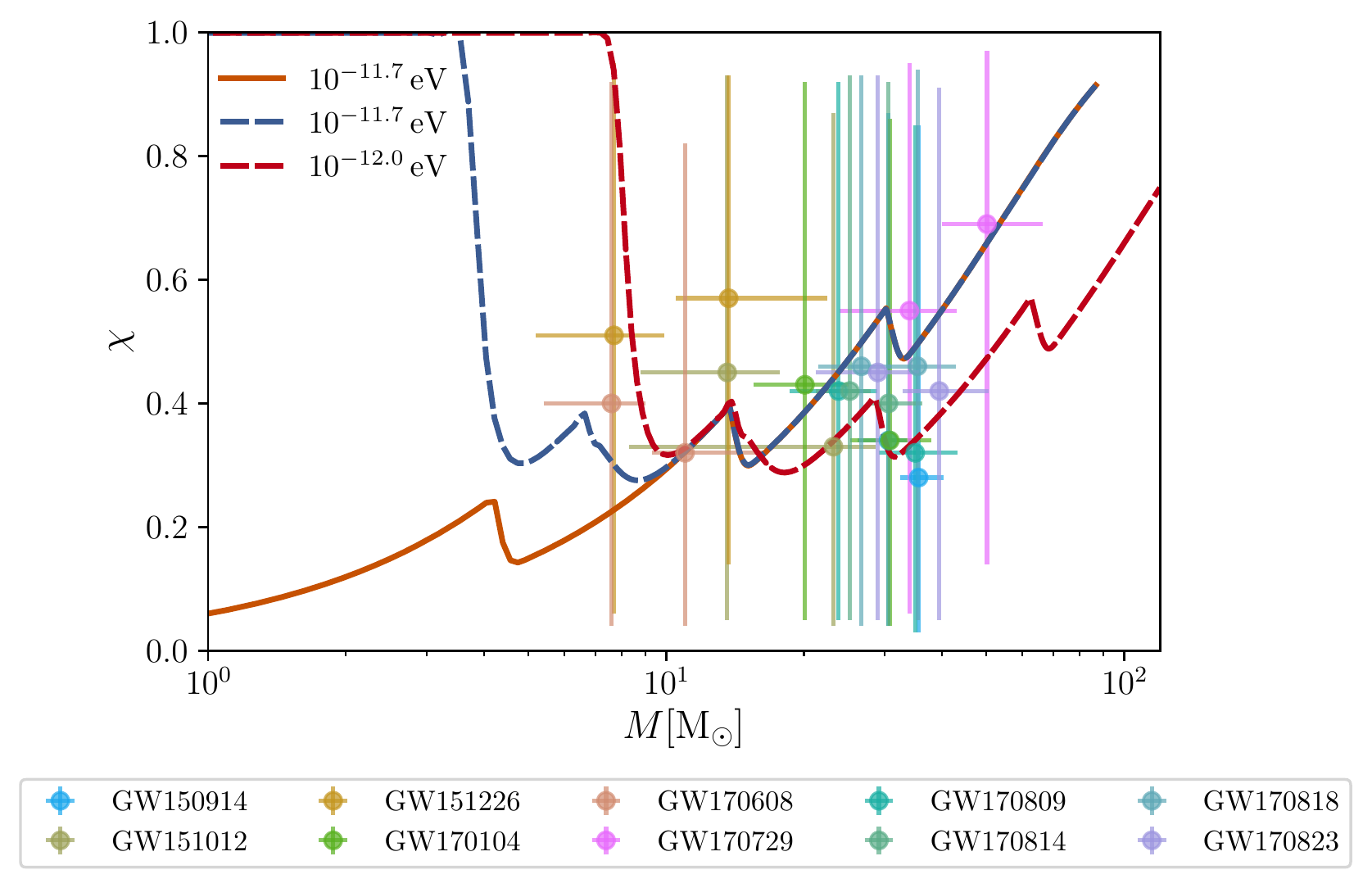}\hfill
\caption{Superradiance exclusion plots superimposed on LIGO spin measurements for the masses that have the largest posterior probability for $\tau_{\rm merger} = 10^{10} {\rm yr}$ for both isolated (solid) and companion (dashed) scenarios. Note that the LIGO spin measurements plotted are $90\%$ confidence intervals using LIGO priors and not using superradiance priors, although they are good visual indicators to estimate which mass of the light boson $\mu$ best explains the data. }
\label{fig:regge}
\end{figure*}

\subsection{Results}
\label{sec:LIGOresults}
The marginal posterior distributions of the boson mass are shown in Fig.~\ref{fig:Posterior_mass} for the isolated (\textit{left}) and companion  (\textit{right}) scenarios. We assume two natal BHs spin magnitude distributions: Flat spin (blue) and Low spin (orange), which we defined in Eq. (\ref{eq:spinprior}). A hypothesis gets the largest support when the superradiance trajectory passes close to the observed values of the spins. This is illustrated in Fig. \ref{fig:regge} for the best fit values of the isolated and companion case for $\tau_{\rm merger} = 10^{10} \, \rm yr$.

For both Flat and Low spin distributions, the posterior of $\mu$ is disfavored for $\mu<10^{-12} \ev$. In this range the superradiance curves are broad enough to shift the prior of the majority of the LIGO events to unacceptably low spins, especially  for GW170729, which is the most massive binary and has the BH with the largest spin.

It is known that LIGO data prefers lower spin values for BHs whose spins are isotropically oriented \cite{LIGOScientific:2018jsj}. The Low spin distribution prior already fits the LIGO data well, hence including superradiance only improves the fit marginally. This can be seen in Fig. \ref{fig:Posterior_mass}, where a broader range of masses for the Low prior fit the data well compared to the Flat prior.

Ultimately, we are interested in testing the superradiance hypothesis against the initial spin prior choice (Flat, Low). For the isolated scenario with $\tau_{\rm merger} = 10^{10} \, \rm{yr}$ (and $\tau_{\rm merger} = 10^{7} \, \rm{yr}$), we find a Bayes factor of $\sim 2.9$ (and $\sim 2.2$) for the case of flat initial spin distribution and $\sim 0.7$ (and $\sim 0.7$) for the case of low initial spin. For the companion case with $\tau_{\rm merger} = 10^{10} \, \rm{yr}$ (and $\tau_{\rm merger} = 10^{7} \, \rm{yr}$) we find a Bayes factor of $\sim 3.8$ (and $\sim 1.2$) for the case of flat initial spin distribution and $\sim 2.8$ (and $\sim 1.2$) for the case of low initial spin. For both isolated and companion scenarios, we find a small preference for superradiance hypothesis. However, given only 20 BH spins, no conclusive, significant statistical statement can be made.

\section{X-ray binaries}
\label{sec:xray}
If the ultralight bosons possess self-interactions, the perturbation created by occupation of the $l$ state prohibits the growth of the $l+1$ level due to level mixing \cite{Arvanitaki_2011}. In this section, we consider the possibility that the BHs in the X-ray binary systems in fact {\em lie on a Regge trajectory}. Our inferred values for the $l$ of the Regge trajectory for each of the X-ray binary BHs are given in Table \ref{table:parameters_xrays}. We parametrize the self-interactions by assuming a pseudoscalar boson with the Lagrangian $$\mathcal{L} \supset \Lambda^{4}\left(1-\cos\left(\frac{\phi}{f_{a}}\right) \right),$$ where $\Lambda$ is the scale associated with the explicit breaking of the $U(1)$ symmetry and $f_{a}$ is the scale associated with spontaneous breaking of the $U(1)$ symmetry. Before we present our results, let us first consider the conditions that need to be satisfied for the consistency of our hypotheses.

If the BH starts with a natal spin above the  $l = q$ Regge trajectory, the $l=q$ superradiance level extracts the spin from the BH until the BH ends on the $l=q$ Regge trajectory. The self-interactions of the bosons occupying the $l=q$ level then act as a {\em perturbation} for the $l=q+1$ state causing level mixing of the superradiant $l=q+1$ and the dumped $l=q-1$ state \cite{Arvanitaki_2011}. This, in turn, causes superradiance to shut down and the BH to stay on the trajectory until the occupation number of $l=q$ level drops through bosons annihilating to gravitons, so that the $l=q+1$ level can proceed to extract the spin. The time spent on the Regge line is given by \cite{Arvanitaki_2015}
\begin{align}
\tau_{\rm Regge} = \left(\rm min(N_{max},N_{bosenova}) \Gamma_{a}\right)^{-1}|\Gamma^{l-1}_{sr}/\Gamma^{l+1}_{sr}|^{1/2}
\end{align}
where $\rm N_{bosenova}$ is the maximum occupation number of the level before the boson cloud collapses in a Bosenova, and $\Gamma_{a}$ is the annihilation rate of the bosons into gravitons, given by \cite{Arvanitaki_2011,Arvanitaki_2015}
\beq
\rm N_{max} &\simeq 10^{76}\left(\frac{\Delta a}{0.1}\right)\left(\frac{M}{10\Msun}\right)^{2}\\
\rm N_{bosenova} &\simeq 5\times 10^{78}\frac{(l+1)^{4}}{\alpha^{3}}\left(\frac{M}{10\Msun}\right)^{2}\left(\frac{f_{a}}{M_{pl}}\right)^{2}\\
\Gamma_{a} &\simeq 10^{-10}\left(\frac{\alpha}{0.5}\right)^{p}\frac{G_{N}}{R^{3}_{g}} \, ,
\eeq
where $p = 17$ for $l=1$ and $p=4l+11$ otherwise.

Larger $f_{a}$ therefore corresponds to lower self-interactions and to shorter times $\tau_{\rm Regge}$ spent on the Regge trajectory: the strongest upper bound on $f_{a}$ which parametrizes the self-interactions of the bosons therefore comes from the oldest X-ray binary (GRS 1915+105) with an age of $4 \, \rm Gyr$. To derive a bound on $f_{a}$, we can therefore use $\tau_{\rm Regge} > 4 \, \rm Gyr $. Assuming that the X-ray binary GRS 1915+105  has been stuck on the $l=q$ trajectory, we get the upperbound $f_a \lessapprox 10^{15}\gev$, fairly independent of $q$. This is consistent with superradiance bounds on self-interactions coming from non-liner effects \cite{Fukuda:2019ewf}. This bound also rules out a QCD axion as an explanation for the observed X-ray binary spins. The reason we consider $q \leq 3$ is because higher levels are not superradiant for  the most massive BH (Cygnus X-1) stuck on $l=q$ \footnote{As we increase $q$ the superradiance levels become closer to each other. Hence, its  possible that the high spin X-ray binaries lie on levels greater than q. This will be taken into account in future work using reasonable astrophysical priors.}. The smaller the $f_{a}$, the longer it takes for a BH to lower its spin to the Regge trajectory, since it has to undergo multiple Bosenovas while reducing its spin. However, since we don't have a prior on the initial spins of the BHs, we don't take this into account.

\begin{figure}
\centering
\includegraphics[width=0.48\textwidth]{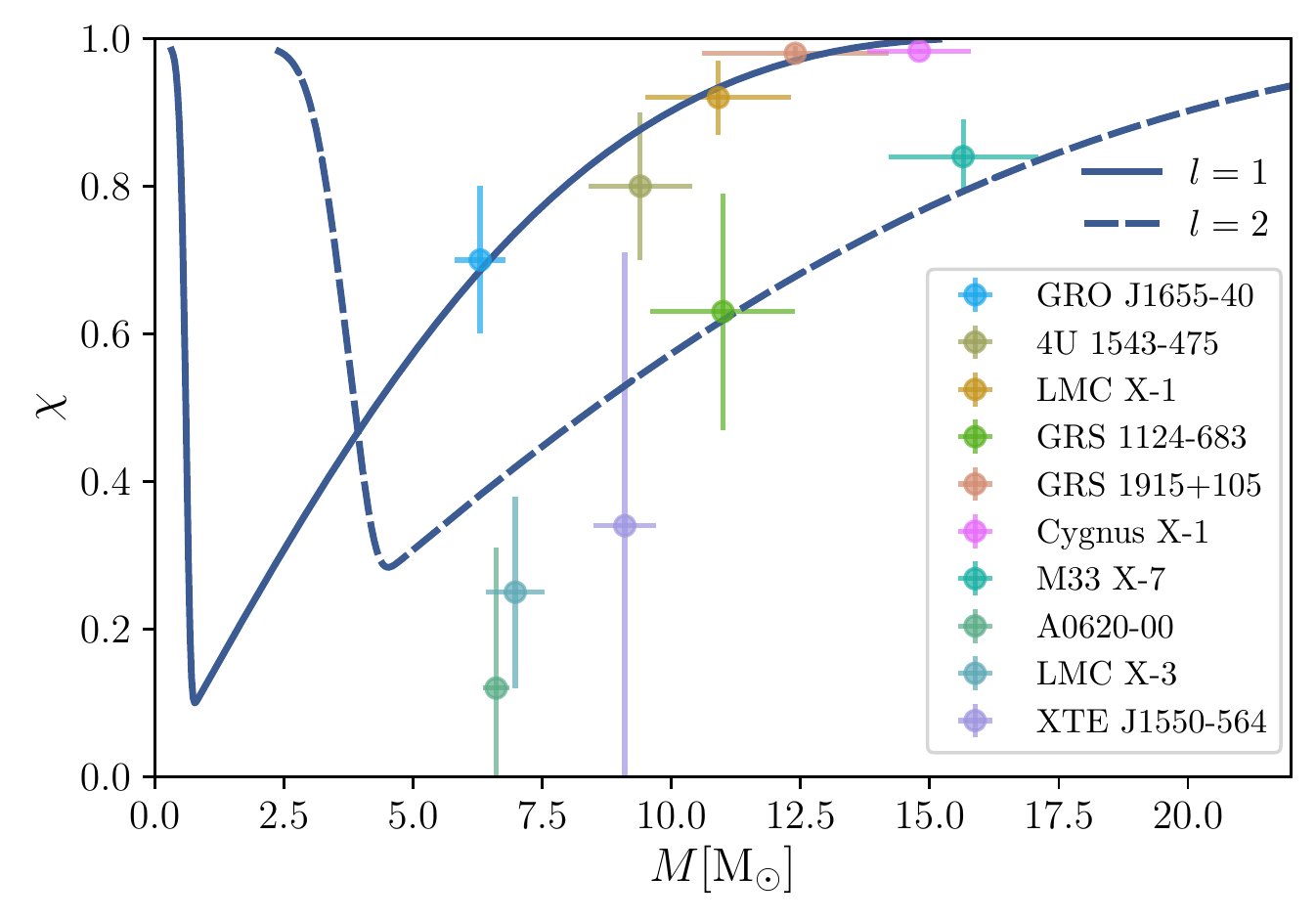}
\caption{Stellar mass BHs from X{-}ray binary systems. The Regge trajectories for $l=1$ (solid) and $l=2$ (dashed) levels for the $\mu = 10^{-11.4} \ev$.}
\label{fig:astro}
\end{figure}

We do a simple $\chi^{2}$ analysis of the mass required to fit the X-ray binary data. The relevant assumptions we make for the mass and spins of the X-ray binary BHs as well as the superradiance levels they are stuck on are given in Table \ref{table:parameters_xrays}. The results for the $95\%$ confidence intervals of mass for a given hypothesis are given in Table \ref{table:results}. All the spins have been calculated using the continuum fitting method (see \cite{McClintock_2013} for a review).

The possibility that X-ray binaries can by themselves corroborate the existence of an ultralight boson is obviously an exciting one. However, a more careful analysis is needed to draw any definitive conclusions. A more thorough analysis of this scenario would involve taking into account the large systematic uncertainties in measuring BH spins\footnote{The spins presented here have used a continuum fitting method. Disagreements exist in the literature regarding the inferred masses and spins of individual BHs using methods other than continuum fitting.}, constructing the correct superradiance priors for arbitrary $f_{a}$, and comparing the likelihood of superradiance being the correct description of the data compared to reasonable astrophysical priors\footnote{The BH in X-ray binaries are believed to have higher spins due to the tidal coupling of its parent star to its binary companion. A proper astrophysical prior will have to take this effect into account.}. We leave this in-depth analysis of X-ray binaries for future work.
\begin{table}
  \centering
\scalebox{0.9}{
\begin{tabular}{ l r r r r r r r r}
\hline 
X-ray Binary & & \multicolumn{1}{c}{$M [\Msun]$}    & &    \multicolumn{1}{c}{$\chi$} & & \multicolumn{1}{c}{Ref} & & \multicolumn{1}{c}{$l$-level}  \\  [0.2ex]  \hline   \hline  \\ [-1.4ex]
GRO J1655-40  & & $6.3^{+ 0.5}_{-0.5}$   & & $0.70^{+0.10}_{-0.10}$  & & \cite{Greene_2001},\cite{Shafee_2005} && $q$ \\ [0.14cm]
4U 1543-475  & & $9.4^{+1.0}_{-1.0}$    & & $0.80^{+0.10}_{-0.10}$  & & \cite{2003IAUS..212.....V},\cite{Shafee_2005} && $q$ \\ [0.14cm]
LMC X-1  & & $10.91^{+1.41}_{-1.41}$    & & $0.92^{+0.05}_{-0.05}$  & & \cite{Orosz_2009},\cite{Gou_2009} && $q$ \\ [0.14cm]
GRS 1124-683  & & $11.0^{+1.4}_{-1.4}$    & & $0.63^{+0.16}_{-0.16}$  & & \cite{Wu_2016},\cite{Chen_2016} && $n$ \\ [0.14cm]
GRS 1915+105  & & $12.4^{+1.8}_{-1.8}$    & & $0.98^{+0.01}_{-0.01}$  & & \cite{Reid_2014},\cite{Miller_2013} && $q$ \\ [0.14cm]
Cygnus X-1  & & $14.8^{+ 1.0}_{-1.0}$   & & \footnote{Note that for Cygnus X-1 only a $3 \sigma$ lower bound on spin is quoted. However since most of the error comes from the mass we assume the spin error to be 0.001 in our analysis. This shouldn't affect our results.}$0.983^{+0.001}_{-0.001}$  & & \cite{Orosz_2011_cyg},\cite{Gou_2014} && $q$ \\ [0.14cm]
M33 X-7  & & $15.65^{+1.45}_{-1.45}$    & & $0.84^{+0.05}_{-0.05}$   & & \cite{Orosz_2007},\cite{Liu_2008} && $n$ \\ [0.14cm]
A0620-00  & & $6.61^{+ 0.25}_{-0.25}$  & & $0.12^{+0.19}_{-0.19}$ & & \cite{Cantrell_2010},\cite{Gou_2010} && N/A \\ [0.14cm]
LMC X-3  & & $6.98^{+ 0.56}_{-0.56}$  & & $0.25^{+0.13}_{-0.13}$  & &  \cite{Orosz_2014},\cite{Steiner_2014} && N/A \\ [0.14cm]
XTE J1550-564  & & $9.10^{+ 0.61}_{-0.61}$     & & $0.34^{+0.37}_{-0.37}$  & & \cite{Orosz_2011},\cite{Steiner_2011} && N/A \\ [0.14cm]\hline
\end{tabular}
}
\caption {Mass and dimensionless spin magnitude of nine X-ray binaries. The level column indicates our hypothesis of the superradiant level the BHs are stuck on. The three BHs with  low spins (A0620-00, LMC X-3, XTE J1550-564) have spins too low to be affected by superradiance. The other BHs are stuck on the Regge trajectory $q$ or $n$ where $q=(1,2,3)$ and $n>q$. In case of asymmetric error bars we conservatively approximate the lower value to be the $1 \sigma$ symmetric errors.}
\label{table:parameters_xrays}
\end{table}

\begin{table}
  \centering
\scalebox{0.9}{
\begin{tabular}{ l c c r r r r r r}
\hline 
$q,n$ & & $\mu$(\rm eV)  \\  [0.2ex]  \hline   \hline  \\ [-1.4ex]
1,2  & & $3.4\times 10^{-12} \leq \mu \leq 5.6\times 10^{-12}$    \\ [0.14cm]
2,4  & & $6.8\times 10^{-12} \leq \mu \leq 1.12\times 10^{-11}$     \\ [0.14cm]
3,5  & & $1\times 10^{-11} \leq \mu \leq 1.63\times 10^{-11}$     \\ [0.14cm]\hline
\end{tabular}
}
\caption {$95\%$ confidence intervals for mass $\mu$ for a given $q,n$. Note that the $n$ has been chosen for a corresponding $q$ to maximize the range of allowed masses.}
\label{table:results}
\end{table}

\section{Summary and Outlook}
\label{sec:conclusion}
Superradiance is a powerful process which might offer a unique channel to detect light scalars. Effects of superradiance on BH spins have been 
theoretically thoroughly investigated. At present, we  have two sources of data for BH spins: binary BH mergers observed by LIGO, and X-ray binaries. In this paper, we discussed the possibility of detecting the presence of a light boson using both LIGO and X-ray binaries. 
Our key findings are as follows:
\begin{itemize}
    \item As shown in Fig. \ref{fig:Posterior_mass}, we can statistically infer the preferred mass of a {\em non-interacting} light bosons that best fits the LIGO data. However, with only 10 detected BH binary mergers, the result is {\em not yet} statistically significant.
    \item The measurement of the intrinsic parameters, especially the spin magnitude, with current second-generation gravitational-wave detectors will remain poorly constrained \cite{Purrer:2015nkh,Vitale:2016avz}, and a third-generation gravitational-wave detector will slightly improve it \cite{Vitale:2016icu}. Only with two third-generation detectors, will we see any significant improvement in the spin magnitude measurements \cite{Vitale:2016icu, Vitale:2018nif}. Therefore, to make any inferences about the BH spins and masses, a hierarchical Bayesian modeling about the BH population is needed. With $\mathcal{O}(1000)$ events expected from the LIGO collaboration in the near future, data from BH binary mergers will be able to constrain or detect {\em non-interacting} light bosons \cite{Ng:2019jsx,LIGOScientific:2018mvr}.
    \item Any analysis that considers the effect of superradiance on LIGO BHs will be very sensitive to the initial BH spin distributions as well as to assumptions about the binary formation process.
    \item X-ray binaries have been used to constrain superradiance \cite{Arvanitaki_2015,Stott_2018,sun2019search,Cardoso_2018}. It is intriguing that the seven measured BHs follow a pattern and give reasonable fit for BHs stuck on Regge trajectories (see Table \ref{table:results}). This is a promising direction. However, a much more careful analysis is needed to make any statistically decisive statements.
\end{itemize}

\section{Acknowledgements}
We thank Colin Bischoff for suggesting we could fit higher $l's$ to the X-ray binary data. We thank Enrico Ramirez-Ruiz for helping us understand X-ray binaries. AG would like to thank the Aspen Center for Physics under NSF grant PHY-1607611 where this work was initiated. AG would like to thank GGI for their hospitality during the completion of part of this work. The research of AG is supported in part by the NSF CAREER grant PHY-1915852. 
NF and SP are partially supported by the U.S. Department of Energy grant number DE-SC0010107. The research of NF is supported by the U.S. Department of Energy grant number DE-SC0017840.


\medskip

\bibliography{references}

\end{document}